
\magnification = 1200
\overfullrule = 0 pt
\baselineskip = 16 pt
\hsize = 14.5 truecm
\vsize = 22. truecm

\font\cp = cmsl9 scaled \magstep0

\def\vac{\vert 0 \rangle}

\def\nis{n_{i \sigma}}
\def\njs{n_{j \sigma}}
\def\aisd{a_{i \sigma}^{\dagger}}
\def\ais{a_{i \sigma}}

\def\ajs{a_{j \sigma}}

\def\niu{n_{i +}}
\def\nju{n_{j +}}
\def\aiud{a_{i +}^{\dagger}}
\def\aiu{a_{i +}}
\def\ajud{a_{j +}^{\dagger}}
\def\aju{a_{j +}}
\def\nid{n_{i -}}
\def\njd{n_{j -}}
\def\aidd{a_{i -}^{\dagger}}
\def\aid{a_{i -}}
\def\ajdd{a_{j -}^{\dagger}}
\def\ajd{a_{j -}}

\def\CcC{{\hbox{\tenrm C\kern-.45em{\vrule height.67em width0.08em depth-.04em
\hskip.45em }}}}
\def\RrR{{\hbox{\tenrm I\kern-.17em{R}}}}
\def\HhH{{\hbox{\tenrm {I\kern-.18em{H}}\kern-.18em{I}}}}
\def\DdD{{\hbox{\tenrm {I\kern-.18em{D}}\kern-.36em {\vrule height.62em
width0.08em depth-.04em\hskip.36em}}}}
\def\ZzZ{{\hbox{\tenrm Z\kern-.31em{Z}}}}
\def\IiI{{\hbox{\tenrm I\kern-.19em{I}}}}
\def\NnN{{\hbox{\tenrm {I\kern-.18em{N}}\kern-.18em{I}}}}

\hrule height 0.0 pt
\nopagenumbers
\line{\hfill POLFIS-TH.22/92}
\vfill
\centerline{\bf THE HALF-FILLED HUBBARD MODEL}
\vskip .15 truein
\centerline{\bf IN THE PAIR APPROXIMATION}
\vskip .15 truein
\centerline{\bf OF THE CLUSTER VARIATION METHOD}
\vskip .4 truein
\centerline{\sl Alessandro Pelizzola}
\vskip 0.5cm

\centerline{Dipartimento di Fisica and Unit\'a INFM,}
\centerline{Politecnico di Torino, I-10129 Torino, Italy}

\vskip .7 truein
\centerline{\bf Abstract}
\vskip .12 truein
{\cp The half filled Hubbard model is studied in the pair approximation of
the Cluster Variation Method. The use of the $SO(4)$ symmetry of the model
makes possible to give a complete analytical characterization of the ground
state, by means of explicit expressions for the double occupancy and the
nearest neighbor correlation functions. The finite temperature analysis is
reduced to the numerical solution of only two coupled transcendental
equations. The behavior of local magnetic moment, specific heat and
correlation functions is given for some typical cases in one and two
dimensions. We obtain good qualitative agreement with exact and numerical
results
in one dimension. The results for finite temperatures show
a rapid evolution, with increasing temperature, from a
strongly antiferromagnetic behavior to a disordered one; in the high
temperature region a maximum (which has been related
to a "gradual" metal--insulator transition) is found in the specific heat
for very large values of the Coulomb repulsion.}
\vfill
\line{P.A.C.S. \# 75.10L \hfill}
\vfill\eject
\pageno 2
\footline{\hss \tenrm -- \folio $\,$ -- \hss}
\hrule height 0 pt
\vskip 2 truecm
{\bf 1. Introduction} \par \medskip
The Hubbard model[1,2] is the simplest model of itinerant electrons which
takes into account the interaction between electrons. It was originally
proposed to describe the behavior of $d$-electrons in transition metals,
and it is expected to describe the metal--insulator (Mott) transition. In the
recent years, the interest about this model has been greatly revived by the
discovery of high--$T_c$ superconductors, since these materials
are generally good Mott insulators and, in the superconducting phase,
exhibit strong antiferromagnetic correlations, just like the half--filled
Hubbard model at low temperatures. \par
The model is defined by the following grand--canonical hamiltonian:
$$H = U \sum_i \niu \nid - \mu \sum_{i \sigma} \nis - t \sum_{\langle i j
\rangle, \sigma} \aisd \ajs, \eqno(1)$$
where $U,t > 0$ and $\ais, \aisd$ and $\nis$ are, respectively, annihilation,
creation
and number operators for electrons at site $i$ with spin $\sigma \in \{ +,-
\}$. The first term represents the Coulomb repulsion between electrons at
the same site (all other interactions are neglected); the second term is
the chemical potential, and the third one is the kinetic term, which
describes hopping of the electrons between sites, with the sum
restricted to non--oriented nearest neighbor (n.n.) pairs. \par
The Hubbard model has been studied by many different
techniques (for reviews see[3,4] and references therein) but an exact
solution is available only in one dimension[5,6], while in two dimensions
or more there are only a few exact results in very particular cases. For
$U/t = 0$ the model describes a
system of non--interacting, moving electrons
and is exactly solvable in any dimension. On the other side, for $U/t =
\infty$ (atomic limit) and at half filling (i.e. $\langle n_i \rangle =
\langle \niu + \nid \rangle = 1$) the ground state is that of an
antiferromagnetic insulator[7], with exactly one electron per site.
At half filling two other very important rigorous results hold:
\item{(i)} the chemical potential is given by $\mu = U/2$ for any value of
$U/t$ and at any temperature[8] and
\item{(ii)} hamiltonian (1) has, for $\mu = U/2$,
an $SO(4)$ symmetry[9]. \par
In this paper we investigate the $D$--dimensional Hubbard model at half
filling in the pair
approximation of the Cluster Variation Method (CVM). The CVM has been
originally introduced by Kikuchi[10] and
its convergence in the thermodynamic limit has been demonstrated by
Schlijper[11]. Recently the method has been given a very elegant
formulation by An[12], in terms of M\"obius inversion. The simplest
level of approximation in the CVM is the site approximation,
which is equivalent to
the ordinary mean--field theory; then we have the pair approximation, which
can be shown to be equivalent to the Bethe approximation. \par
The pair approximation of the CVM has already been applied to the Hubbard
model in refs.[13-16]. Unfortunately, in
these references only the $U(1) \otimes U(1)$ Cartan subgroup
of the $SO(4)$ symmetry group later studied
by Yang and Zhang was used, and the authors had to deal with large sets of
coupled transcendental equations, which they could solve only for relatively
high
temperatures ($kT/t > 1$, with $k$ Boltzmann's
constant and T absolute temperature). Furthermore, in [14-16]
equivalence is assumed between sites belonging to the two interpenetrating
sublattices which form a bipartite lattice. \par
In this paper we apply the pair approximation of the CVM to the Hubbard
model on a bipartite lattice (that is, a lattice which is made of two
interpenetrating sublattices, say $A$ and $B$, in such a way that a site
belonging to
sublattice $A$ has all its nearest neighbors in the $B$ sublattice and
viceversa: examples are the 1-D chain, hc, sq, sc and bcc lattices)
by taking into account the full $SO(4)$ symmetry of the model. Furthermore, the
equivalence between the two sublattices is not assumed, but it is derived
from the thermodynamics of the model.
The ground state is determined analytically, i.e.
explicit expressions are derived for the double occupancy and the n.n.
correlation functions at $T = 0$, for any number of dimensions and any
value of the interaction $U/t$. For the finite temperature case, the
problem is reduced to the numerical solution of two coupled transcendental
equations. Such equations can be solved at any
temperature and in the limit $T \to 0$ the ground state solution is
recovered. \par
The validity of the approximation is first checked by comparing the
behavior of the
zero temperature local magnetic moment vs. $U/t$ in $D = 1$ with the exact
solution
for the infinite chain reported by Hirsch[17], and then by
comparing the same quantity at finite temperature for typical values of
$U/t$ with the numerical results for finite chains obtained by Shiba and
Pincus[18]. Once the validity of the method has been established, we report
on the numerical results at finite temperature for $D \ge 1$: correlation
functions, hopping expectations and specific heat are given in some typical
cases,
and the antiferromagnetic behavior of the system as well as the inhibition of
certain hopping processes at low temperatures are discussed. Finally, it is
noticed that a high temperature maximum appears in the specific heat
for very large values of the interaction,
in agreement with previous studies where this maximum was related to a
gradual metal--insulator transition. \par
The paper is organized as follows: in Sec. 2 we construct the trial free
energy according to the pair approximation of the CVM, taking into
account the symmetry of the hamiltonian. In Sec. 3 the ground state is
obtained and discussed, and the zero temperature local magnetic
moment is compared with the exact result for the infinite chain. In Sec. 4
the analysis is extended to finite temperature and the behavior of
various physical quantities is given and discussed and, finally, in Sec. 5
some conclusions are drawn. \par \bigskip

{\bf 2. Free energy} \par \medskip
Following An's formulation[12], the CVM trial free energy for a bipartite
lattice in the pair approximation can be written as
$$f = {E \over N} + k_B T\left\{ {1 - z \over 2} \left[
{\rm Tr}(\rho_A \ln \rho_A) + {\rm Tr}(\rho_B \ln \rho_B) \right] +
{z \over 2} {\rm Tr} (\rho_p \ln \rho_p) \right\}, \eqno(2)$$
where $E$ is the internal energy, $N$ is the number of lattice sites,
and $\rho_A$, $\rho_B$ and $\rho_p$ are the reduced density matrices
(to be determined by minimizing $f$) for a site
belonging to sublattice $A$, a site belonging to sublattice $B$ and a pair
of nearest neighbors, respectively. \par
Before taking the variation of $f$ with respect to the reduced density
matrices, let us determine which constraints for such density matrices can
be derived from the symmetry group of the hamiltonian.
As shown in [9], hamiltonian (1), with $\mu = U/2$ because of the half
filling condition, commutes with a $SO(4) = \displaystyle{SU(2) \otimes
SU(2) \over \ZzZ_2}$ group, where one $SU(2)$ (referred to as
the magnetic one) is generated by
$$J_z = {1 \over 2}\sum_i (\niu - \nid), \quad
J_+ = \sum_i \aiud \aid, \quad
J_- = \sum_i \aidd \aiu \eqno(5)$$
the other $SU(2)$ (called the pairing, or superconductive one), which is
relevant only at half--filling, by
$$K_z = {1 \over 2}\sum_i (\niu + \nid - 1), \quad
K_+ = \sum_i e^{i\phi_i} \aiud \aidd, \quad
K_- = \sum_i e^{i\phi_i} \aid \aiu \eqno(6)$$
(the phase factor $e^{i\phi_i}$ is $+1$ for sites in sublattice $A$ and
$-1$ for sites in sublattice $B$) and $\ZzZ_2$ interchanges the two $SU(2)$
symmetries. The presence or absence of this symmetry in the quantum state
of the system characterizes the
different phases: a disordered phase will be invariant under
the whole $SO(4)$ symmetry group, while a phase with magnetic and/or
superconductive order will be invariant under a reduced symmetry group, with
both $SU(2)$ (or one) spontaneously broken down.
Since in this paper we are concerned with the half filled case,
and it is conjectured that at half filling the Hubbard model does not
undergo any phase transition, we
devote our attention to the disordered phase,
thus assuming the whole $SO(4)$ symmetry (in view of the good agreement
with exact results in one dimension, this assumption should be correct also
at zero temperature). The possibility of a
phase transition, associated with a spontaneous breaking of the magnetic
$SU(2)$
symmetry group, will be examined in a forthcoming paper[19] for the extended
Hubbard model at general filling. \par
In order to impose the commutation relations between the reduced density
matrices and the $SO(4)$ generators defined above, we introduce in the site and
pair reduced Fock spaces the customary basis of eigenstates of the number
operators. In such a
basis, requiring that the reduced density matrices commute with the Cartan
operators $J_z$ and $K_z$, $\rho_A$ and $\rho_B$
turn out to be diagonal, while for $\rho_p$ one obtains the same block
structure as in [16], with only 36 non--zero elements. By imposing
furthermore the commutation with $J_+$ and $K_+$ (or, equivalently, with their
hermitian conjugates $J_-$ and $K_-$), one finds
that $\rho_\gamma$ ($\gamma = A,B$) has two distinct eigenvalues, say
$d_\gamma$ and $\displaystyle{1 \over 2} - d_\gamma$, each with
multiplicity 2, whereas $\rho_p$ is a block diagonal matrix with:
\item{i)} two eigenvalues $\lambda_1$ and $\lambda_2$, each with
multiplicity 2;
\item{ii)} four degenerate $2 \times 2$ blocks, which give rise to two
eigenvalues $\lambda_3$ and $\lambda_4$ with multiplicity 4;
\item{iii)} a $4 \times 4$ block with eigenvalues $\lambda_1, \lambda_2,
\lambda_5$ and $\lambda_6$. \par
Summarizing, we have 6 different
eigenvalues, $\lambda_1$ and $\lambda_2$ with multiplicity $m_1 = m_2 = 3$,
$\lambda_3$ and $\lambda_4$ with multiplicity $m_3 = m_4 = 4$,
and $\lambda_5$ and $\lambda_6$ with multiplicity $m_5 = m_6 = 1$.
Of these, only 5 are independent, because of the normalization condition
${\rm Tr}(\rho_p) = 1$. \par
Recalling that the expectation value of an
operator $X$ is given by $\langle X \rangle = {\rm Tr}(\rho X)$, one can
compute, with some simple algebra, the expectation values of all the site
and n.n. pair operators. The non--zero expectation values turn out to be
($i$ and $j$ nearest neighbors, $i \in A$ and $j \in B$)
$$\eqalign{
& \langle \nis \rangle = \langle \njs \rangle = {1 \over 2} \cr
& \langle \niu \nid \rangle = d_A \cr
& \langle \nju \njd \rangle = d_B \cr
& c_p \equiv \langle \niu \nju \rangle = \langle \nid \njd \rangle =
\lambda_1 + \lambda_2 + \lambda_3 + \lambda_4 \cr
& c_a \equiv \langle \niu \njd \rangle = \langle \nju \nid \rangle =
\lambda_1 - \lambda_2 + {1 - d_A - d_B \over 2} \cr
& \langle \niu \nid \nju \rangle = \langle \niu \nid \njd \rangle =
{1 \over 2} \left( d_A + c_p + c_a - {1 \over 2} \right) \cr
& \langle \nju \njd \niu \rangle = \langle \nju \njd \nid \rangle =
{1 \over 2} \left( d_B + c_p + c_a - {1 \over 2} \right) \cr
& q \equiv \langle \niu \nid \nju \njd \rangle = \lambda_1 \cr} \eqno(7)$$
for the diagonal operators (notice that in this scheme the half filling
condition is derived from the symmetry and not imposed), and
$$\eqalign{
& p \equiv \langle \aiud \aidd \ajd \aju \rangle = {1 \over 2} - c_p - c_a \cr
& p^\prime \equiv \langle \aidd \ajud \aiu \ajd \rangle = c_a - c_p \cr
& {\tau_0 \over 2} \equiv
\langle \aisd \ajs n_{i -\sigma} n_{j -\sigma} \rangle =
\langle \aisd \ajs (1 - n_{i -\sigma}) (1 - n_{j -\sigma}) \rangle \cr
& \qquad = {1 \over 2} \sqrt{(\lambda_3 - \lambda_4)^2 - {1 \over 4}(d_A -
d_B)^2} \cr
& {\tau_1 \over 2} \equiv
\langle \aisd \ajs (1 - n_{i -\sigma}) n_{j - \sigma} \rangle
= \langle \aisd \ajs n_{i -\sigma} (1 - n_{j -\sigma}) \rangle \cr
& \qquad = {1 \over 4}\sqrt{(\lambda_5 - \lambda_6)^2 - \left[ 2(d_A + d_B) - 1
-
3(\lambda_1 - \lambda_2) \right]^2} \cr} \eqno(8)$$
(together with the obvious hermitian conjugates) for the non--diagonal
operators. \par
The free energy per site, as a function of $d_A$, $d_B$ and $\lambda_i, i =
1, \ldots 6$, is then
$$\eqalign{
f = \quad & {U \over 2}(d_A + d_B - 1) - 2zt \sqrt{(\lambda_3 - \lambda_4)^2 -
{1 \over 4}(d_A - d_B)^2} \cr
& -zt \sqrt{(\lambda_5 - \lambda_6)^2 - \left[ 2(d_A + d_B) - 1 -
3(\lambda_1 - \lambda_2) \right]^2} \cr
& + kT (1 - z) \sum_{\gamma = A,B}
\left[ d_\gamma \ln d_\gamma + \left( {1 \over 2} - d_\gamma \right)
\ln \left( {1 \over 2} - d_\gamma \right) \right] \cr
& + kT{z \over 2} \sum_{i=1}^6 \left( m_i \lambda_i \ln \lambda_i \right). \cr}
\eqno(9)$$
\par \bigskip

{\bf 3. The ground state} \par \medskip
At $T = 0$, the free energy per site is but the internal energy, and is
given by
$$\eqalign{
f = \quad & {U \over 2}(d_A + d_B - 1) - 2zt \sqrt{(\lambda_3 - \lambda_4)^2 -
{1 \over 4}(d_A - d_B)^2} \cr
& -zt \sqrt{(\lambda_5 - \lambda_6)^2 - \left[ 2(d_A + d_B) - 1 -
3(\lambda_1 - \lambda_2) \right]^2}. \cr} \eqno(10)$$
The ground state can thus be obtained by minimizing $f$ with respect to the
$d_\gamma$ and the $\lambda_i$. Since we are
looking for an absolute minimum and our variables are subject to
constraints (the eigenvalues of the density matrices, as well as the
arguments of the square roots in (10) must be non--negative, and
$\rho_p$ must be properly normalized),
we should search our minimum possibly at the domain boundary of the
constrained variables, and not only in the interior. \par
Indeed, the minimum is found for
$$d_A = d_B \equiv d = {1 \over 4} \left( 1 - {{\cal U} \over \sqrt{{\cal U}^2
+ 16}} \right),
\qquad {\cal U} = {U \over zt}, \eqno(11)$$
$\lambda_5 = 1$ and $\lambda_i = 0, i \ne 5$. \par
The ground state is thus described by
$$\eqalign{
& d_A = d_B = d = {1 \over 4} \left( 1 - {{\cal U} \over
\sqrt{{\cal U}^2 + 16}} \right) \cr
& c_a = {1 \over 2} - d = {1 \over 4} \left( 1 + {{\cal U} \over
\sqrt{{\cal U}^2 + 16}} \right) \cr
& \tau_1 = \sqrt{2d (1 - 2d)} = {2 \over \sqrt{{\cal U}^2 + 16}} \cr
& q = c_p = \tau_0 = 0. \cr} \eqno(12)$$ \par
The configuration of a pair of nearest neighbors in the ground
state can be derived as the eigenvector of $\rho_p$ corresponding to the
eigenvalue $\lambda_5 = 1$. One obtains, up to a normalization constant,
$$\eqalign{ \vert \Psi \rangle = \quad & \left[ {2 \over \sqrt{{\cal U}^2 +
16}}
(\aiud \aidd + \ajud \ajdd) \right. \cr
& + \left. {1 \over 2} \left( 1 + {{\cal U} \over
\sqrt{{\cal U}^2 + 16}} \right) (\aiud \ajdd + \ajud \aidd) \right] \vac. \cr}
\eqno(13)$$ \par
It is worth noticing that such a configuration is the superposition of an
antiferromagnetic singlet pair (equivalent to that used by Anderson[20] in
the construction of his RVB state) with a fraction of doubly occupied
sites. The concentration of doubly occupied sites as well as the kinetic
energy (the expectation value of the hopping term) vanish for large
${\cal U}$ and have their maximum for ${\cal U} = 0$. The n.n. correlations are
strictly antiferromagnetic (i.e., $c_p = 0$), as expected, and no phase
transition is found in any number of dimensions.
Some hopping processes, e.g. the hopping of an
electron from a doubly occupied to a singly occupied site, or from a singly
occupied to an empty site, turn out to be inhibited in the ground state. \par
In order to check the validity of our approximation, we compare the
local magnetic moment at $T = 0$ vs. $U/t$ for the
1-D chain ($z = 2$) with the exact result reported in [18]. The local
magnetic moment $S$ is proportional to the expectation value of the square of
the magnetization:
$$S = {3 \over 4} \langle (\niu - \nid)^2 \rangle \eqno(14)$$
and is directly related to the double occupancy, since
$$S = {3 \over 4} (1 - 2d) = {3 \over 8} \left( 1 + {{\cal U} \over
\sqrt{{\cal U}^2 + 16}} \right). \eqno(15)$$
(15) is exact both in the non--interacting case (${\cal U} = 0, S = 3/8$)
and in the atomic limit (${\cal U} = \infty, S = 3/4$).
Fig.~1 shows the comparison between our results for $z = 2$ (solid line) and
the exact solution for the 1-D chain (circles). The agreement is within
10\% for all values of $U/t$.
\par\bigskip
{\bf 4. Finite temperature} \par \medskip
Since we have found $d_A = d_B = d$ in the ground state, and the entropy
contribution favors this latter condition, one can expect this symmetry
relation to
hold even at finite temperature. Furthermore, a breaking of such symmetry
at finite temperature would yield a reentrant phase with staggered double
occupancy, and there is no indication of such phases in the Hubbard model.
Indeed, we have checked numerically that the minima of $f$ always appear
for $d_A = d_B = d$. We shall therefore assume from now on the latter
relation. \par
In this way we obtain a free energy which is a function of
six independent variables only: $d$ and five of the $\lambda_i$. Instead of
minimizing $f$ directly, we introduce the following new set of
indipendent variables:
$$\eqalign{
& \delta = 4d - 1 \cr
& r_{n,n+1} = m_n (\lambda_n - \lambda_{n+1}), \qquad n = 1,3,5 \cr
& R_{n,n+1} = m_n (\lambda_n + \lambda_{n+1}), \qquad n = 1,3, \cr}\eqno(16)$$
and we define $R_{56} = 1 - R_{12} - R_{34}$. After rewriting $f$, as given
by (2), in terms of the above variables, the minimum-$f$ requirement
gives (assuming, with no loss of generality, $\lambda_3 > \lambda_4$)
$$\eqalign{
& 0 = {\partial f \over \partial \delta} = {U \over 4} + zt {\delta -
r_{12} \over 2 \tau_1} + kT {1 - z \over 2} \ln {1 + \delta \over 1 -
\delta} \cr
& 0 = {\partial f \over \partial r_{12}} = -zt {\delta - r_{12} \over 2
\tau_1} + kT {z \over 4} \ln {R_{12} + r_{12} \over R_{12} - r_{12}} \cr
& 0 = {\partial f \over \partial r_{34}} = - {z \over 2} t + kT {z \over 4}
\ln {R_{34} + r_{34} \over R_{34} - r_{34}} \cr
& 0 = {\partial f \over \partial r_{56}} = - {zt \over 2 \tau_1} r_{56} +
kT {z \over 4} \ln {R_{56} + r_{56} \over R_{56} - r_{56}} \cr
& 0 = {\partial f \over \partial R_{12}} = kT{z \over 4} \left(
\ln {R_{12}^2 - r_{12}^2 \over 36} - \ln {R_{56}^2 - r_{56}^2 \over 4}
\right) \cr
& 0 = {\partial f \over \partial R_{34}} = kT{z \over 4} \left(
\ln {R_{34}^2 - r_{34}^2 \over 64} - \ln {R_{56}^2 - r_{56}^2 \over 4}
\right) . \cr}\eqno(17)$$
Upon defining $x = r_{12} - \delta$ and after some algebra three of the above
equations can be solved for $r_{34}, R_{12}, R_{34}$ and $\delta$ (or $r_{12}$)
leaving us with the following two coupled transcendental
equations for $x$ and $r_{56}$:
$$\eqalign{
& x = \tanh \left[ {z \over 2(z - 1)} \beta \left( {x \over \tau_1} -
{{\cal U} \over 2} \right) \right] - 6R \sinh \left( \beta {x \over \tau_1}
\right) \cr
& r_{56} = 2R \sinh \left( \beta {r_{56} \over \tau_1} \right), \cr}
\eqno(18)$$
where $\beta = (kT/t)^{-1}$, $\tau_1 = \displaystyle{{1 \over 2}
\sqrt{r_{56}^2 - x^2}}$ and
$$R = \left[ 6 \cosh \left( \beta {x \over \tau_1} \right) + 8 \cosh \beta +
2 \cosh \left( \beta {r_{56} \over \tau_1} \right) \right]^{-1}. \eqno(19)$$
Once (18) has been solved, the remaining variables are given by the
following relations:
$$\eqalign{
& r_{12} = -6R \sinh \left( \beta {x \over \tau_1} \right) \cr
& R_{12} = 6R \cosh \left( \beta {x \over \tau_1} \right) \cr
& r_{34} = 8R \sinh \beta \cr
& R_{34} = 8R \cosh \beta. \cr} \eqno(20)$$ \par
As a check for the whole procedure, we compare in Fig.~2 our results for the
local
magnetic moment for $z = 2$ and some typical values of $U/t$ with the
exact (numerical) results obtained by Shiba and Pincus[18] for a six sites
chain with periodic boundary conditions. We find good qualitative agreement,
and again differences are contained within 10\%.
It can be observed that the solution for low
temperatures converges to the value predicted by the ground state analysis.
The results for the chain are compared with those for the square lattice
($z = 4$) in
Fig.~3: our analysis shows, as expected from numerical simulation[17],
that the local moment increases
with increasing $U/t$ and with decreasing dimensionality. \par
In Fig.~4 we report the correlation functions $c_p$ (lower curves) and
$c_a$ (upper curves), in Fig.~5 the hopping contributions $\tau_0$ (lower
curves) and $\tau_1$ (upper curves), in Fig.~6 the "double hoppings" $p$
(lower curves) and $p^\prime$ (upper curves) and in Fig.~7 the specific heat,
for
$U/t = 8$ (solid lines) and $U/t = 4$ (dashed lines) for a square lattice. \par
Of course there is no evidence of a true phase transition (the specific heat
exhibits a maximum but not a sharp peak), but we can clearly distinguish
a low temperature behavior ($kT/t < 0.5$) from a high temperature one ($kT/t
> 1$). The low temperature region is characterized by strong
antiferromagnetic correlations and by a relatively large kinetic energy
associated to the moving electrons, due almost entirely to double hoppings
and to hopping processes from doubly occupied to empty sites and viceversa,
while the remaining processes are
strongly inhibited because of the ground state configuration.
The high temperature region, besides, looks like a true
disordered phase, with almost equally distributed correlations ($c_p
\approx c_a \approx 1/4$) and low kinetic energy. Furthermore in this
region, as already noticed in [16], for very large values of the
interaction $U/t$ a spread maximum appears
in the specific heat, which was related by Ho and Barry to a "gradual"
metal--insulator transition. \par \bigskip
{\bf 5. Conclusions} \par \medskip
We have investigated the half--filled Hubbard model in the pair
approximation of the Cluster Variation Method, making use of the full
$SO(4)$ symmetry of the model. We have given an analytical description of the
ground state, by means of the double occupancy and of the n.n. correlation
functions and, for finite temperature, we have derived a pair of coupled
transcendental equations. Numerical solution shows two different
behaviors, connected by a smooth but rapid change in the values of the
parameters. The low temperature behavior is strongly antiferromagnetic and
exhibits the inhibition of certain hopping processes, while a large kinetic
energy is associated with the others. In the high temperature region we find a
quite disordered behavior, with a spread maximum, which was related to a
metal--insulator transition, for very large values of the interaction. Good
agreement is found with exact and numerical results in one dimension. \par
\bigskip
{\bf Acknowledgements} \par
The author is grateful to Prof. M. Rasetti and to Dr.
A. Montorsi for many stimulating discussions and careful reading of the
manuscript. \vfill \eject

{\bf References}
\item{[1]} M.C. Gutzwiller, Phys. Rev. Lett. {\bf 10} (1963) 159.
\item{[2]} J. Hubbard, Proc. Roy. Soc. {\bf A276} (1963) 238.
\item{[3]} A. Montorsi (ed.), {\it The Hubbard Model -- A Reprint Volume},
World Scientific, Singapore, 1992.
\item{[4]} M. Rasetti (ed.), {\it The Hubbard Model -- Recent Results},
World Scientific, Singapore, 1992.
\item{[5]} E.H. Lieb and F.Y. Wu, Phys. Rev. Lett. {\bf 20} (1968) 1445.
\item{[6]} B.S. Shastry, Phys. Rev. Lett. {\bf 56} (1986) 2453.
\item{[7]} P.W. Anderson, Phys. Rev. {\bf 115} (1959) 2.
\item{[8]} D.C. Mattis and L.F. Landovitz, J. Non--Cryst. Solids {\bf 2}
(1970) 454.
\item{[9]} C.N. Yang and S.C. Zhang, Mod. Phys. Lett. {\bf B4} (1990) 759.
\item{[10]} R. Kikuchi, Phys. Rev. {\bf 81} (1951) 988.
\item{[11]} A.G. Schlijper, Phys. Rev. {\bf B27} (1983) 6841.
\item{[12]} G. An, J. Stat. Phys. {\bf 52} (1988) 727.
\item{[13]} C.C. Chen, Phys. Rev. {\bf B16} (1977) 1312.
\item{[14]} T. Ogawa, T. Ogawa and K.A. Chao, Phys. Rev. {\bf B17} (1978) 4124.
\item{[15]} C.C. Chen and M. H. Huang, J. Appl. Phys. {\bf 50} (1979) 1761.
\item{[16]} W.C. Ho and J.H. Barry, Phys. Rev. {\bf B20} (1979) 2118.
\item{[17]} J.E. Hirsch, Phys. Rev. {\bf B 22} (1980) 5259.
\item{[18]} H. Shiba and P. Pincus, Phys. Rev. {\bf B5} (1972) 1966.
\item{[19]} A. Pelizzola, in preparation.
\item{[20]} P.W. Anderson, Science {\bf 235} (1987) 1196; for further
references to RVB see A.P. Balachandran, E. Ercolessi, G. Morandi and A.M.
Srivastava, Int. J. Mod. Phys. {\bf B4} (1990) 2057.
\vfill\eject
{\bf Figure Captions} \par \bigskip
\parindent .4 truein
\item{}
\itemitem{\hbox to .67 truein{Fig. 1: \hfill}} Local magnetic moment at $T
= 0$. Our result (solid line) and result from [17] (circles).
\itemitem{\hbox to .67 truein{Fig. 2: \hfill}} Local magnetic moment at
finite temperature for $U/t = 8$ (upper curves) and $U/t = 4$ (lower
curves). Solid lines are our results, circles are from [18].
\itemitem{\hbox to .67 truein{Fig. 3: \hfill}} Local magnetic moment at
finite temperature for $U/t = 8$ (upper curves) and $U/t = 4$ (lower
curves). Dashed lines are for the linear chain and solid lines for the
square lattice.
\itemitem{\hbox to .67 truein{Fig. 4: \hfill}} Correlation functions $c_p$
(lower curves) and $c_a$ (upper curves) on the square lattice for $U/t = 8$
(solid lines) and
$U/t = 4$ (dashed lines).
\itemitem{\hbox to .67 truein{Fig. 5: \hfill}} Hopping expectation values
$\tau_0$ (lower curves) and $\tau_1$ (upper curves) on the square lattice for
$U/t = 8$ (solid
lines) and $U/t = 4$.
\itemitem{\hbox to .67 truein{Fig. 6: \hfill}} Double hoppings $p$ (lower
curves) and $p^\prime$ (upper curves) on the square lattice for $U/t = 8$
(solid lines) and
$U/t = 4$ (dashed lines).
\itemitem{\hbox to .67 truein{Fig. 7: \hfill}} Specific heat on the square
lattice for $U/t = 8$ (solid lines) and $U/t = 4$.
\vfill\eject\end